\documentclass[prl,twocolumn,showpacs,amsmath,amssymb]{revtex4-1}

\usepackage{graphicx}% Include figure files
\usepackage{subfigure}
\usepackage{dcolumn}% Align table columns on decimal point
\usepackage{array}
\usepackage{bm}% bold math
\usepackage[hang]{footmisc}

\usepackage[usenames,dvipsnames]{color}

\begin{document}

\title{\bf{Antiferroelectricity and ferroelectricity in epitaxially strained PbZrO$_3$ from first principles} \\[11pt] } \author{Sebastian E. Reyes-Lillo and Karin M. Rabe}
\affiliation{Department of Physics and Astronomy, Rutgers University, Piscataway, NJ 08854-8019}\date{\today}

\begin{abstract}
Density functional calculations are performed to study the effect of epitaxial strain on PbZrO$_3$. We find a remarkably small energy difference between the epitaxially strained polar \textsl{R3c} and nonpolar \textsl{Pbam} structures over the full range of experimentally accessible epitaxial strains -3~$\%$~$\leqslant$~$\eta$~$\leqslant$~4~$\%$. While ferroelectricity is favored for all compressive strains, for tensile strains the small energy difference between the nonpolar ground state and the alternative polar phase yields a robust antiferroelectric ground state. The coexistence of ferroelectricity and antiferroelectricity observed in thin films is attributed to a combination of strain and depolarization field effects.
\end{abstract}


\maketitle

%%%%%%%%%%%%%%%%%%%%%%%%%%%%%%%%%%%
\marginparwidth 2.5in
\marginparsep 0.5in
\def\jab#1{\marginpar{\small JAB: #1}}
\def\jwb#1{\marginpar{\small JWB: #1}}
\def\hit#1{\marginpar{\small HIT: #1}}
\def\amr#1{\marginpar{\small AMR: #1}}
\def\scr{\scriptsize}
%%%%%%%%%%%%%%%%%%%%%%%%%%%%%%%%%%%

%%%%%%%%INTRODUCTION--------------------------------------------------------------------

There is a renewed interest in antiferroelectric materials driven by potential technological applications. An antiferroelectric~\cite{Rabe2013} is like a ferroelectric~\cite{Lines, Rabe} in that its structure is obtained through distortion of a nonpolar high-symmetry reference structure; for ferroelectrics the distortion is polar, while for antiferroelectrics it is nonpolar. However, not all nonpolar phases thus obtained are antiferroelectric; in addition, there must be an alternative ferroelectric phase obtained by a polar distortion of the same reference structure, close enough in free energy so that an applied electric field can induce a first-order phase transition from the antiferroelectric to the ferroelectric phase, producing a characteristic polarization-electric field (P-E) double-hysteresis loop. The electric-field-induced transition is the source of functional properties and promising technological applications. Non-linear strain and dielectric responses at the phase switching are useful for transducers and electro-optic applications~\cite{Berlincourt1966, Zhang2009}. The shape of the double hysteresis loop suggests applications in high-energy storage capacitors~\cite{Jaffe1961, Love1990}. In addition, an effective electro-caloric effect can also be induced in systems with a large entropy change between the two phases~\cite{Mischenko2006}. 
\medskip

Lead zirconate PbZrO$_3$ (PZO) was the first material identified as antiferroelectric~\cite{Shirane1951}. Despite extensive studies and characterization, PZO continues to offer insights into the origin and complexity of antiferroelectricity~\cite{Liu2011, Tan2011}. In bulk form, PZO has a cubic perovskite structure at high temperatures and a nonpolar orthorhombic ground state below $T_c~\sim$~505~K. The ground state has space group \textsl{Pbam}~\cite{Fujishita1982,Fujishita1984} and
unit cell dimensions $\sqrt{2}a_0 \times 2\sqrt{2}a_0 \times 2a_0$ with respect to the reference lattice constant $a_0$. Its distorted perovskite structure is derived from the cubic (C) unit cell through a nonpolar $\Sigma_2$ distortion mode of Pb$^{+2}$ ion displacements in the $\langle 110 \rangle_C$ direction, combined with oxygen octahedron rotation $R^-_5$ modes around the $\langle 110 \rangle_C$ axis ($a^-a^-c^0$ in Glazer notation). Under an applied electric field, PZO single crystals undergo a first order phase transition into a sequence of polar phases with rhombohedral symmetry~\cite{Fesenko1978}. Similar rhombohedral polar phases are observed in the polycrystalline ceramic system Pb(Zr$_{1-x}$Ti$_{x}$)O$_3$ under small 5-10~$\%$ isovalent substitution of zirconium for titanium~\cite{Sawaguchi1953, Viehland1995}. In thin films, the competition between the rhombohedral low-energy structures and the PZO ground state is less studied. Room temperature ferroelectricity have been reported below a certain critical thickness~\cite{Ayyub1998} and under large compressive epitaxial strain~\cite{Chaudhuri2011}. Under different circumstances, large remnant polarizations~P$_r$~$\sim$~5-20~$\mu$C/cm$^2$ have been measured in P-E antiferroelectric-like double hysteresis loops~\cite{Yamakawa1996, Pintilie2008, Chaudhuri2011, Liu2012}, suggesting coexistence of ferroelectricity with antiferroelectricity.
\medskip

In this paper, we present first principles calculations performed to investigate the effect of epitaxial strain on the structure and stability of PZO. In bulk, we find a small energy difference of~$\sim$~1~meV/f.u between the nonpolar ground state \textsl{Pbam} and the alternative polar phase \textsl{R3c}. Under epitaxial strain, a small energy difference between these two competing low-energy phases persists over a remarkably wide range of experimentally accessible epitaxial strain. While ferroelectricity is favored at compressive strains, the nonpolar ground state is favored at tensile strains. In the strain regime where the nonpolar phase is lower in energy, the small energy difference between the nonpolar and polar phases ensures antiferroelectricity. The coexistence of ferroelectricity and antiferroelectricity observed in thin films is attributed to a combination of strain and depolarization field effects. 
\medskip

%%%%%%%%%%%%%%FIRST-PRINCIPLES DETAILS------------------------------------------------------------

We performed density-functional theory (DFT) calculations using version 6.4.1~of~\texttt{ABINIT}~\cite{Gonze2009} package. The local-density approximation~(LDA), a plane-wave energy cutoff of 680~eV, and a $4 \times 4 \times 4$ Monkhorst-Pack sampling of the Brillouin zone~\cite{Monkhorst1976} were used for all structural optimizations. Polarization was calculated in a $10 \times 10 \times 10$ grid using the modern theory of polarization~\cite{King-Smith1993} as implemented in \texttt{ABINIT}. We used norm-conserving pseudopotentials from the Bennett-Rappe library~\cite{Bennett2012} with reference configurations: \textsl{Pb}([\textsl{Hg}]6p$^0$), \textsl{Zr}([\textsl{Kr}]4d$^0$5s$^0$) and \textsl{O}(1s$^2$2s$^2$2$p^4$), generated by the \texttt{OPIUM} code~\cite{opium}.  In order to allow direct comparison with experiments, the epitaxial strain diagram was constructed with respect to $a_0=4.1$~$\AA$, which is the cube root of the calculated volume per f.u. of the \textsl{Pbam} ground state. This reference lattice constant coincides with the optimized lattice constant of the cubic perovskite structure, and underestimate the experimental value of~4.16~$\AA$~\cite{Zhang2011}~by~1.5~$\%$.
\medskip
 
\begin{figure}[t]
\includegraphics[scale=0.50]{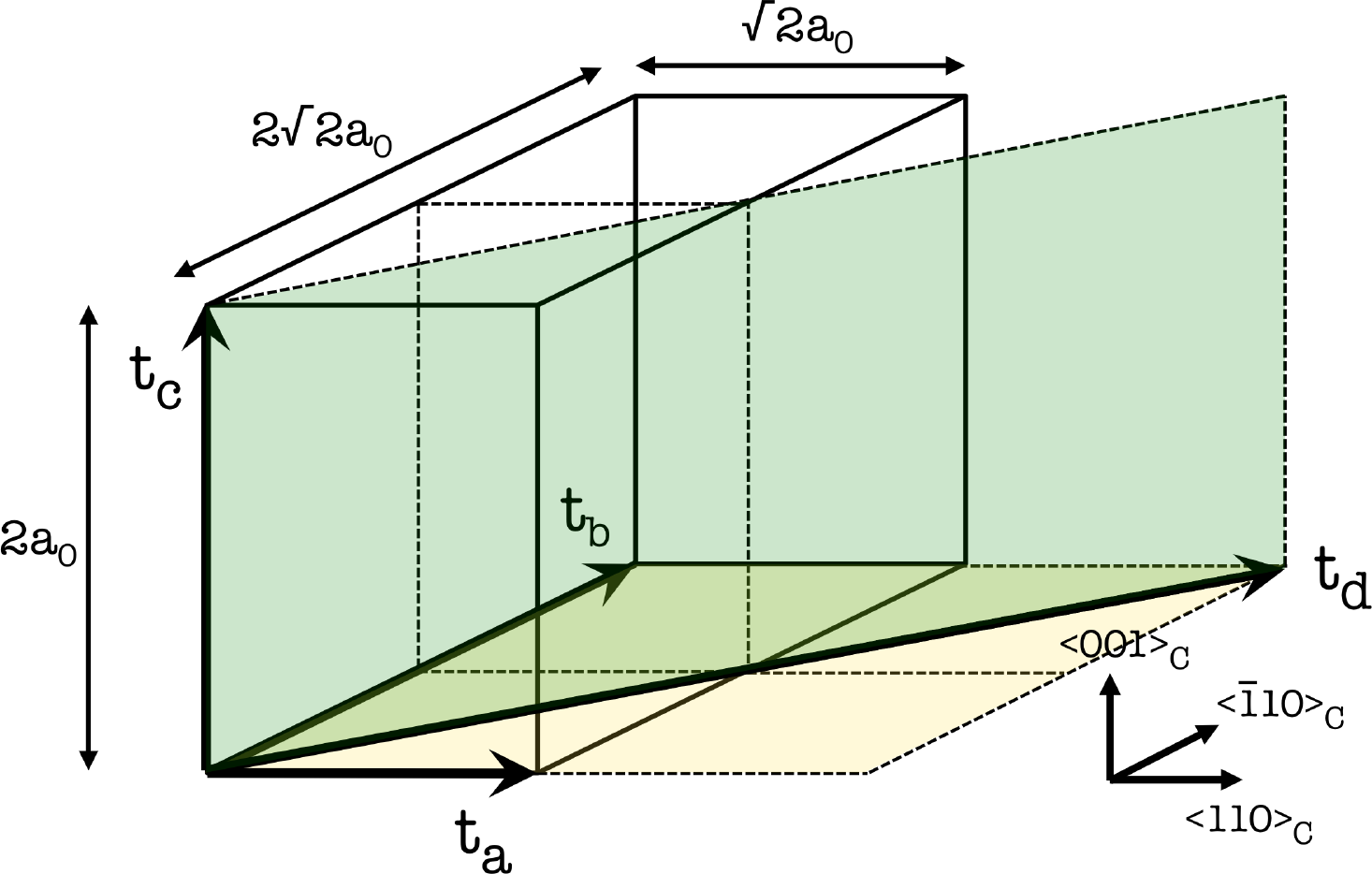}
\caption{\label{fig:matchingplanes.pbam} 
Lattice vectors of the \textsl{Pbam} ground state structure. The $\sqrt{2}a_0 \times 2\sqrt{2}a_0 \times 2a_0$ unit cell is shown with solid lines. While the (001) plane is defined by the lattice vectors $\texttt{t$_a$}$ and $\texttt{t$_b$}$, the (120) plane is defined by the lattice vectors $\texttt{t$_c$}$ and $\texttt{t$_d$}$.
}
\includegraphics[scale=0.50]{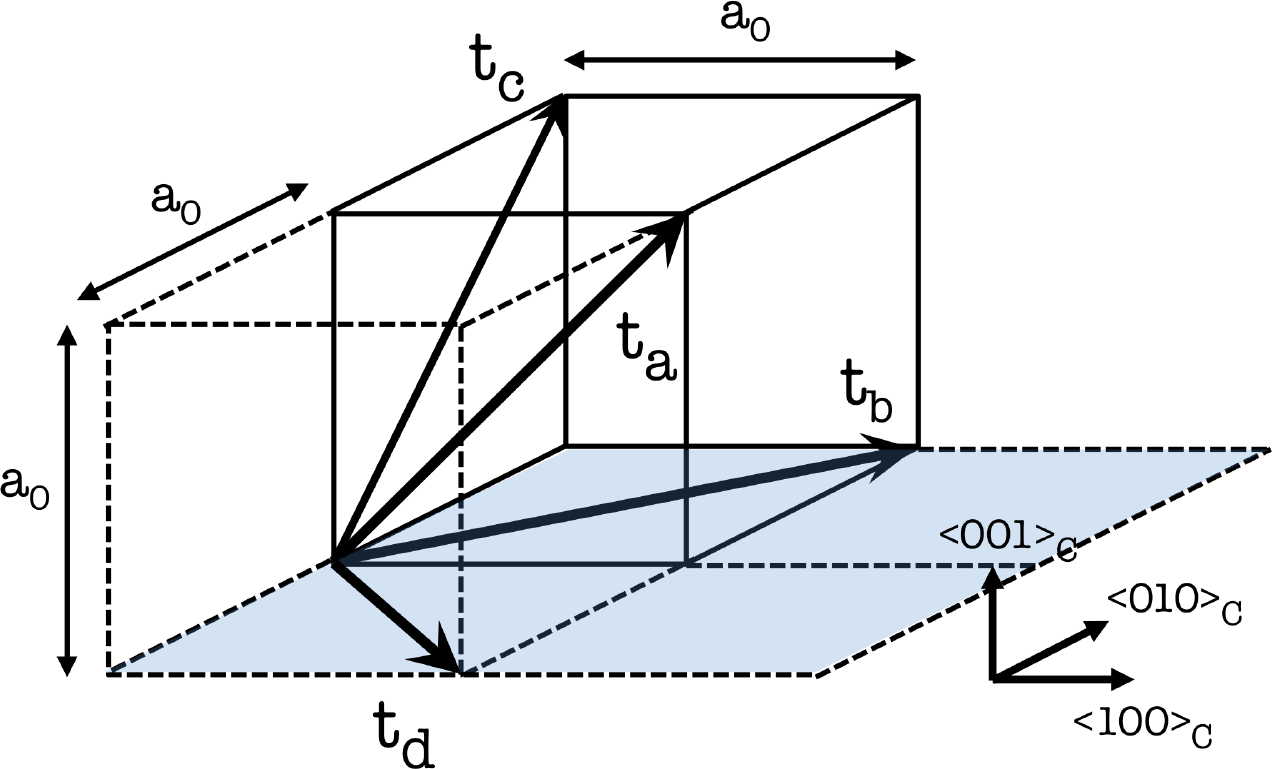}
\caption{\label{fig:matchingplanes.r3c} 
Lattice vectors of the \textsl{R3c} phase. The epitaxial plane (001) is defined by the lattice vectors $\texttt{t$_b$}$ and $\texttt{t$_d$}$.
}
\end{figure}

The effect of epitaxial strain was investigated through ``strained-bulk" calculations~\cite{Pertsev1988, Dieguez2005}. As shown in Fig.~\ref{fig:matchingplanes.pbam}, the unit cell of the \textsl{Pbam} ground state structure contains two symmetry-inequivalent primitive perovskite planes, (001) and (120) ((001)$_C$ and (010)$_C$ with respect to the cubic perovskite vectors), and therefore allows two distinct orientations for epitaxial growth over a square terminated (001)$_C$ perovskite substrate. Epitaxial strain is imposed on the structure by fixing the two lattice vectors defining the matching plane and optimizing the length and direction of the third, out-of-plane, lattice vector, along with the atom positions, until the forces on the atoms are less than 0.05~meV/A. We designate epitaxially strained phases as \textsl{ePbam} to distinguish them from bulk \textsl{Pbam}. We stress that while the space group is preserved as \textsl{Pbam} when (001) is chosen as the matching plane (\textsl{c-ePbam}), the symmetry is lowered to \textsl{P2/m} when the (120) plane is chosen as the matching plane (\textsl{ab-ePbam}). Similarly, the symmetry of the rhombohedral phase \textsl{R3c} is lowered to monoclinic \textsl{Cc} by epitaxial strain on a square substrate, and we refer to this phase as \textsl{eR3c} in the rest of the paper. For epitaxial strain calculations, we map the \textsl{eR3c} structure into the unit cell defined by \{$\texttt{t$_{a}$}$,$\texttt{t$_{b}$}$,$\texttt{t$_{d}$}$\} in Fig.~\ref{fig:matchingplanes.r3c}, where the lattice vectors $\texttt{t$_b$}$ and $\texttt{t$_d$}$ define the (001) matching plane.
\medskip

%%%%%%%%%%%%%%FIRST-PRINCIPLES RESULTS------------------------------------------------------------
We begin by identifying low-energy bulk PZO structures obtained by distortion of the cubic perovskite structure through unstable modes, as calculated in the phonon spectrum by linear response DFT~\cite{Ghosez1999}. We focus our attention on polar and nonpolar structures generated by the unstable polar mode and by unstable oxygen octahedron rotation modes. In close analogy with PbTiO$_3$, the strong instability of PZO at $\Gamma$ is a result of the well-known lone-pair stereochemical activity of Pb atoms. In the absence of zone boundary modes, freezing-in the polar $\Gamma_4^-$ mode leads to ferroelectric phases similar to BaTiO$_3$, see Table~\ref{table:polar-phases}. In the hypothetical \textsl{P4mm} structure, the polar instability induces a large displacement $\sim$~0.65~$\AA$ of Pb$^{+2}$ ions against the oxygen octahedron network and a large polarization, comparable to PbTiO$_3$. The addition of out-of-phase octahedron rotations along the $\langle 111 \rangle_C$ direction of \textsl{R3m} ($c^{-}c^{-}c^{-}$ in Glazer notation) leads to the metastable \textsl{R3c} phase, with LiNbO$_3$ structure type. The small energy difference between this phase and the \textsl{Pbam} ground state suggests this structure as the most promising candidate for the field induced ferroelectric phase~\cite{Tan2011}.
\medskip

\begin{table}[t]
\caption{Space group, energy gain $\Delta E$ (meV/f.u.), polarization magnitude P ($\mu$C/cm$^2$), estimated equilibrium strain for the (001)$_C$ ($\sigma_c$) and (100)$_C$ ($\sigma_a$) matching planes, and volume expansion $\Delta V/V$ ($\%$) of selected polar structures.}
\begin{ruledtabular}
\begin{tabular}{lccccc}
Space group &$\Delta E$ &P &$\sigma_c$ &$\sigma_{a}$ &$\Delta V/V$ \\ \hline
\begin{tabular}{l}  \textsl{Pm$\bar{3}$m}\\ \textsl{P4mm}\\ \textsl{Amm2}\\ \textsl{R3m} \\ \end{tabular}
&\begin{tabular}{c}  0\\ 248\\ 272\\ 299 \\ \end{tabular}
&\begin{tabular}{c}  0\\ 78 \\ 77 \\ 80 \\ \end{tabular}
&\begin{tabular}{c}  \phantom{-}0\\ -0.74\\ \phantom{-}1.59\\ \phantom{-}0.94\\ \end{tabular}
&\begin{tabular}{c}  \phantom{-}0\\ \phantom{-}1.43\\ \phantom{-}0.45\\ \phantom{-}0.94\\ \end{tabular}
&\begin{tabular}{c}  0\\ 2.07\\ 2.50\\ 2.83\\ \end{tabular} \\ \hline 
\begin{tabular}{l}  \textsl{R3c}\\ \end{tabular}
&\begin{tabular}{c}  344\\ \end{tabular}
&\begin{tabular}{c}  102\\ \end{tabular}
&\begin{tabular}{c}  \phantom{-}0.14\\ \end{tabular}
&\begin{tabular}{c}  \phantom{-}-\\ \end{tabular}
&\begin{tabular}{c}  0.51\\ \end{tabular}
\end{tabular}
\end{ruledtabular}
\label{table:polar-phases}
\end{table}

The strong rotational instability of PZO produces low-energy structures even in the absence of polar distortions. As shown in Table~\ref{table:rotational-phases}, coupling between $R$ and $M$ point octahedron rotations can further decrease the energy of the system by inducing displacement of Pb atoms. The symmetry lowering from combinations of oxygen octahedron rotation modes can induce additional zone-boundary distortions, such as the $X_5^-$ mode in \textsl{Pnma} ($a^{+}b^{-}b^{-}$) and \textsl{P4$_2$/nmc} ($a^{+}a^{+}c^{-}$), and the $R_4^-$ mode in \textsl{Cmcm} ($a^{0}b^{+}c^{-}$).
\medskip

Structures obtained by freezing-in selected additional unstable zone-boundary modes are reported in Table~\ref{table:nonpolar-phases}. In the absence of octahedron rotations, the polar \textsl{R3m} structure has a lower energy than the four f.u. \textsl{Pbam} structure (see Table~\ref{table:polar-phases} and~\ref{table:nonpolar-phases}). While the addition of out-of-phase octahedron rotations in the \textsl{R3m} structure leads to the \textsl{R3c} phase with $\sim$~45~meV/f.u. energy gain, addition of $a^{-}a^{-}c^{0}$ octahedron rotations in the four~f.u. \textsl{Pbam} structure produces the observed eight~f.u. \textsl{Pbam} structure with $\sim$~65~meV/f.u. energy gain, favoring the nonpolar ground state. 
\medskip

\begin{table}[t]
\caption{Space group, formula units (f.u.), Glazer notation, energy gain $\Delta E$ (meV/f.u.), estimated equilibrium strain for the (001)$_C$ ($\sigma_c$) and (010)$_C$ ($\sigma_{b}$) matching planes, and volume expansion $\Delta V/V$ ($\%$) of $15$ possible~\cite{Woodward2005} combinations of $M$ and $R$ point rotation modes.}
\begin{ruledtabular}
\begin{tabular}{lcccccc}
Space group & f.u. & Glazer not. & $\Delta E$  & $\sigma_c$ & $\sigma_{b}$ & $\Delta V/V$ \\ \hline
\begin{tabular}{l}  \textsl{Pm$\bar{3}$m}\\ \textsl{Im$\bar{3}$} \\ \textsl{R$\bar{3}$c}\\ \textsl{I4/mcm}\\ \textsl{I4/mmm}\\ \textsl{P4$_2$/nmc}\\ \textsl{P4/mbm}\\ \textsl{Imma}\\ \textsl{Immm}\\  \textsl{Cmcm}\\ \textsl{Pnma}\\ \textsl{C2/c}\\ \textsl{C2/m}\\ \textsl{P2$_1$/m}\\ \textsl{P$\bar{1}$}\\ \end{tabular}
&\begin{tabular}{c}  1\\ 8\\ 2\\ 4\\ 8\\ 8\\ 2\\ 4\\ 8\\  8\\ 4\\ 4\\ 4\\ 4\\ 4\\ \end{tabular}

&\begin{tabular}{l}  $a^{0}a^{0}a^{0}$\\ $a^{+}a^{+}a^{+}$\\ $c^{-}c^{-}c^{-}$\\ $a^{0}a^{0}c^{-}$\\ $a^{0}b^{+}b^{+}$\\ $a^{+}a^{+}c^{-}$\\ $a^{0}a^{0}c^{+}$\\ $a^{0}b^{-}b^{-}$\\ $a^{+}b^{+}c^{+}$\\ $a^{0}b^{+}c^{-}$\\ $a^{+}b^{-}b^{-}$\\ $a^{-}b^{-}b^{-}$\\ $a^{0}b^{-}c^{-}$\\  $a^{+}b^{-}c^{-}$\\ $a^{-}b^{-}c^{-}$\\\end{tabular}
&\begin{tabular}{c}   0\\   212\\   270\\   234\\   216\\   276\\    214\\   291  \\  $^{\ast}$\textsl{Im$\bar{3}$}\\ 282\\   306 \\ $^{\ast}$\textsl{Imma}\\ $^{\ast}$\textsl{Imma}\\ $^{\ast}$\textsl{Pnma}\\ $^{\ast}$\textsl{Imma}\\ \end{tabular}
&\begin{tabular}{c}  \phantom{-}0\\  -0.24\\   -0.39\\  -1.36\\  \phantom{-}0.08\\  -0.68\\  -1.38\\  \phantom{-}0.09  \\  \\  -0.66\\  -0.45 \\ \\ \\ \\ \\ \end{tabular}
&\begin{tabular}{c}  \phantom{-}0\\  -0.24\\   -  \\  \phantom{-}0.02\\  -0.47\\   -0.21\\  \phantom{-}0.10\\  -0.64  \\  \\ -0.61\\  -0.49 \\ \\ \\ \\ \\ \end{tabular}
&\begin{tabular}{c}  \phantom{-}0\\ -0.61\\ -1.07\\ -1.25\\ -0.77\\  -0.99\\ -1.11\\ -1.09  \\  \\ -1.12\\ -1.32 \\ \\ \\ \\ \\ \end{tabular}
\end{tabular}
\end{ruledtabular}
\footnotesize{$\ast$ convergence to this high-symmetry structure}
\label{table:rotational-phases}
\end{table}

\begin{table}[t]
\caption{Space group, formula units (f.u.), relevant mode content, energy gain $\Delta E$ (meV/f.u.), estimated equilibrium strain for the (001)$_C$ ($\sigma_c$) and (010)$_C$ ($\sigma_{ab}$) matching planes, and volume expansion $\Delta V/V$ ($\%$) of selected nonpolar structures.}
\begin{ruledtabular}
\begin{tabular}{lcccccc}
Space group  & f.u. & mode  & $\Delta E$ & $\sigma_{c}$  & $\sigma_{ab}$ & $\Delta V/V$   \\ \hline
\begin{tabular}{l}   \textsl{Cmcm}\\ \textsl{Pmma} \\ \textsl{Pbam}\\ \end{tabular}
&\begin{tabular}{c}  2\\ 2\\ 4 \\ \end{tabular}
&\begin{tabular}{c}  $X_5^-$\\ $M_5^-$\\ $\Sigma_2$-$M_5^-$\\ \end{tabular}
&\begin{tabular}{c}  71\\ 147\\ 288 \\ \end{tabular}
&\begin{tabular}{c}  \phantom{-}0.72\\ \phantom{-}1.49\\ \phantom{-}1.56\\ \end{tabular}
&\begin{tabular}{c}  \phantom{-}0.71\\ -0.35\\ \phantom{-}0.35\\ \end{tabular}
&\begin{tabular}{c}  2.22\\ 2.08\\ 2.37\\ \end{tabular} \\ \hline
\begin{tabular}{l}  \textsl{Pbam}\\ \end{tabular}
&\begin{tabular}{c}  8\\ \end{tabular}
&\begin{tabular}{c}  $\Sigma_2$-$R_5^-$\\ \end{tabular}
&\begin{tabular}{c}  345\\ \end{tabular}
&\begin{tabular}{c}  \phantom{-}0.62\\ \end{tabular}
&\begin{tabular}{c}  -0.33\\ \end{tabular}
&\begin{tabular}{c}  0.05\\ \end{tabular}
\end{tabular}
\end{ruledtabular}
\label{table:nonpolar-phases}
\end{table}

Fully relaxation of the \textsl{Pbam} and \textsl{R3c} structures leads to a remarkably small energy difference of $\sim$~1~meV/f.u. between them. The volume of the \textsl{R3c} polar structure is slightly larger (0.47~\%) than that of the nonpolar \textsl{Pbam} structure. The experimental lattice constants of \textsl{Pbam}, a~$=$~5.8736~$\AA$, b~$=$~11.7770~$\AA$ and c~$=$~8.1909~$\AA$, at 10~K ~\cite{Fujishita2003} are underestimated by the calculated lattice constants, a~$=$~5.8253~$\AA$, b~$=$~11.7199~$\AA$ and c~$=$~8.1072~$\AA$, by 1~$\%$, typical of LDA. For comparison, we also calculated the energy difference using the experimental \textsl{Pbam} volume and found that the energy difference between the observed \textsl{Pbam} and the hypothetical field-induced \textsl{R3c} structure is also $\sim$~1~meV/f.u., comparable to previous results~\cite{Singh1995, Johannes2005, Kagimura2008}. 
\medskip

%%%%%%%%%%%%%% EPITAXIAL STRAIN------------------------------------------------------------
Next, we consider the effect of epitaxial strain on the relative stability of various structures. The degree of stabilization of a certain structure can be determined by comparing the shape and dimensions of its relaxed structure with the epitaxial strain conditions. For a given structure, of relaxed unit cell lattice vectors \{$\texttt{t$_{a}$}$,$\texttt{t$_{b}$}$,$\texttt{t$_{c}$}$\}, and a given matching plane, with out-of-plane lattice vector $\texttt{t$_{j}$}$, we estimate the equilibrium energy minimum at $\sigma_{j}=100 \times (1/2) \times \sum_{i} (\texttt{t$_i$}-\texttt{t$_{i0}$})/\texttt{t$_{i0}$}$ epitaxial strain; where $i$ denote the two lattice vectors defining the matching plane, and the reference lattice vectors \{$\texttt{t$_{a0}$}$,$\texttt{t$_{b0}$}$,$\texttt{t$_{c0}$}$\} are the relevant linear combination of the cubic perovskite vectors with $a_0=4.1$~$\AA$. As an example, the relaxed lattice parameters of the \textsl{Pbam} ground state are compared with the corresponding edges of the $\sqrt{2}a_0 \times 2\sqrt{2}a_0 \times 2a_0$ unit cell (see Fig.~\ref{fig:matchingplanes.pbam}). The results are in Table~\ref{table:nonpolar-phases}; while $\sigma_c$ estimates the energy minimum of \textsl{c-ePbam} at~$\sim$~0.62~$\%$~tensile strain, $\sigma_{ab}$ estimates the energy minimum of \textsl{ab-ePbam} at~$\sim$~0.33~$\%$~compressive strain. This illustrates how this approach can help identify phases that would be favored by nonzero epitaxial strain.
\medskip 

\begin{figure}[t]
\includegraphics[scale=0.45]{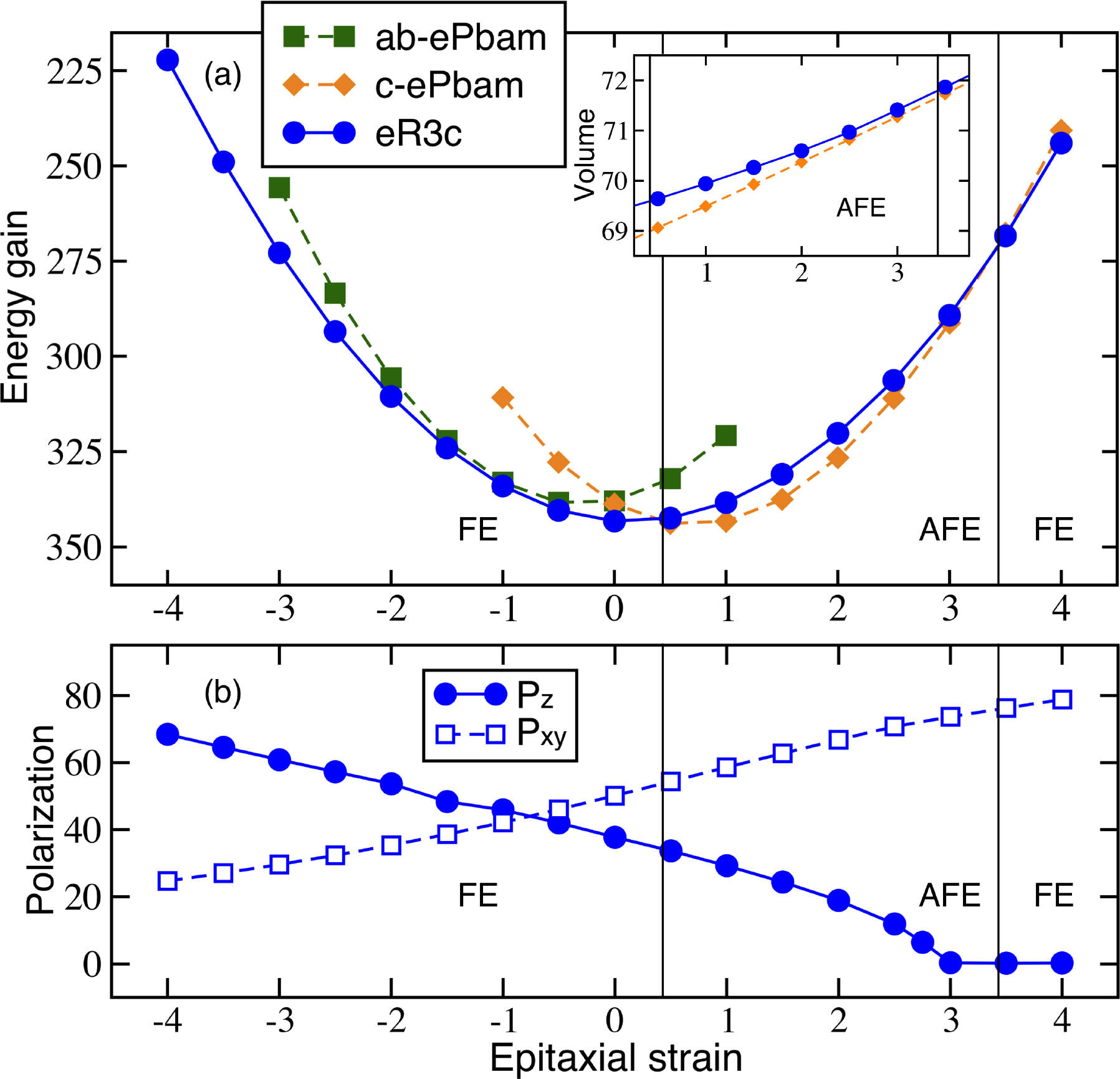}
\caption{\label{fig:epitaxial} 
(a) Energy (meV/f.u.)~vs~epitaxial strain~($\%$) diagram. Epitaxial strain is computed as described in the text. FE and AFE refers here to ferroelectric and antiferroelectric ground state, respectively. Inset: Volume per f.u. ($\AA^3$)~vs~epitaxial strain~($\%$) in the AFE region. (b) Polarization ($\mu$C/cm$^2$) components of the \textsl{eR3c} phase as a function of epitaxial strain ($\%$). P$_z$ and P$_{xy}$ denote the perpendicular and parallel components with respect to the matching plane.}
\end{figure}

The bulk energies of the calculated structures and their $\sigma$ values for the relevant matching planes are shown in Table~\ref{table:polar-phases},~\ref{table:rotational-phases} and~\ref{table:nonpolar-phases}. Based on the assumption that these phases have comparable effective elastic constants to those of \textsl{Pbam} and \textsl{R3c}, we note that the energy gain $\Delta E$ at the optimal equilibrium strain $\sigma$ of these structures is not large enough to overcome the energy cost relative to \textsl{Pbam} and \textsl{R3c}. Through this simple argument, we conclude therefore that they will not be stabilized at $|\eta|~<$~4~\% epitaxial strain; this has been verified for the case of \textsl{Pnma}. The effects of epitaxial strain on \textsl{R3c} and \textsl{Pbam}, the lowest-energy structures of PZO, are shown in Fig.~\ref{fig:epitaxial}~(a). The relaxed structure of \textsl{Pbam} has a large contraction of~$\sim$~1.27~$\%$ in the $\texttt{t$_c$}$ axis and a large expansion of~$\sim$~0.3~$\%$ and ~$\sim$~0.9~$\%$ in the $\texttt{t$_a$}$ and $\texttt{t$_b$}$ axis, explaining the large separation between energy minima of \textsl{ePbam} (large value of $\sigma_c-\sigma_{ab}$ as discussed in the previous paragraph) and the robust ground state at tensile strain. Around 0~$\%$ strain, the in-plane lattice constants of the \textsl{Pbam} structure are less compatible with the square-lattice epitaxial constraint, and the elastic energy costs of deforming the bulk equilibrium state lift the \textsl{c-ePbam} energy curve above the energy curve of the \textsl{eR3c} phase. Phonon eigenfrequencies of \textsl{c-Pbam} calculated at selected values of~1~and~3~$\%$ tensile strain show no further instabilities and confirm its stability against polar distortions.
\medskip

We now focus on the remarkably small energy difference between \textsl{ePbam} and \textsl{eR3c} over nearly the entire range of strain. While the ferroelectric \textsl{eR3c} phase is favored for strains less than~0.4~$\%$, between~0.4~$\%$ and~3.4~$\%$ tensile strain, the \textsl{c-ePbam} phase is favored over the \textsl{eR3c} phase. Throughout this range, the energy difference between the nonpolar and polar structures is smaller than~$\sim$~7~meV/f.u., leading to antiferroelectricity. The ferroelectric \textsl{eR3c} phase is again stabilized between~3.4~$\%$ and~5~$\%$ strain, while the four f.u. \textsl{Pbam} structure is the lowest energy state above 5~$\%$. While within the accuracy of our calculations it is not possible precisely to predict the critical strains that will be observed in experiments, we expect semiquantitative agreement. In the region where antiferroelectricity is stabilized, the antiferroelectric-ferroelectric field induced transition between the \textsl{c-ePbam} ground state and the \textsl{eR3c} phase has a maximum volume expansion of~$\sim$~0.85~$\%$ at~$\sim$~0.4~$\%$ tensile strain (see inset of Fig.~\ref{fig:epitaxial}~(a)). The effect of epitaxial strain on the polarization of the \textsl{eR3c} phase is shown in Fig.~\ref{fig:epitaxial}~(b).  
\medskip

%%%%%%%%DISCUSSION--------------------------------------------------------------------
The computed epitaxial strain diagram can be used to interpret P-E hysteresis loops observed in PZO films. Films under tensile strain exhibit classic double hysteresis loops~\cite{Boldyreva2007}, consistent with our results. Under compressive strain, antiferroelectric-like double loops with non-zero remnant polarization have been observed~\cite{Boldyreva2007, Pintilie2008, Chaudhuri2011, Liu2012} with a magnitude P$_r$ proportional to the film conductivity. This apparent inconsistency with our results can be resolved by recognizing that observation of ferroelectricity in a film requires compensation of the depolarization field. In highly insulating high-quality films, with negligible compensation of the depolarization field, electrostatic energy would suppress the ferroelectric phase in favor of a nonpolar antiferroelectric phase~\cite{Boldyreva2007}. In samples with free carriers available to compensate the depolarization field, nonzero remnant polarization would arise from the ferroelectric phase present in the coherently strained region near the interface~\cite{Chaudhuri2011}. In highly coherent thin films, compensation of the depolarization field would favor ferroelectric behavior. Finally, in thick films, relaxation of the majority of the film to the bulk antiferroelectric \textsl{Pbam} phase can also account for the observed double loops. 
\medskip

%%%%%%%%CONCLUSION--------------------------------------------------------------------
In summary, two different structures of PZO, one nonpolar \textsl{ePbam} and one polar \textsl{eR3c}, are very close in energy and compete under the effect of epitaxial strain. While ferroelectricity is stabilized at compressive epitaxial strain, antiferroelectricity is favored at tensile strains.

\section{\label{sec:level1} Acknowledgements}

We would like to thank J. Bennett, K. Garrity, O. Dieguez and D. Vanderbilt for useful discussion. This work was supported by the Office of Naval Research Grant No.~N00014-12-1-1040. S.E.R.-L. would also like to thank the support of Conicyt and the sponsor of Fulbright Foundation.

\end{document}